# Pseudo-ϵ-Expansion and the Two-Dimensional Ising Model

## A. I. Sokolov

*St. Petersburg State Electrotechnical University, ul. Professora Popova 5, St. Petersburg, 197376 Russia*
*e-mail: ais2002@mail.ru*

**Abstract**—The pseudo-ϵ-expansions for the coordinate of the fixed point $g^*$, the critical exponents, and the sextic effective coupling constant $g_6$ are determined for the two-dimensional Ising model on the basis of the five-loop renormalization group series. It is found that the pseudo-ϵ-expansions for the coordinate of the fixed point $g^*$, the inverse exponent $\gamma^{-1}$, and the constant $g_6$ possess a remarkable property, namely, the higher terms of these series are so small that reliable numerical results can be obtained without invoking Borel summation.

## 1. INTRODUCTION

The two-dimensional Ising model, for which Onsager obtained the exact solution 60 years ago, has been widely used as a proving ground for testing approximate methods [1–10]. Recently [7, 10], the five-loop contributions to the renormalization group (RG) functions were determined for a theoretical-field version of this model, namely, the two-dimensional scalar Euclidean theory $\lambda\varphi^4$. These results, together with known four-loop expansions [4], made it possible to obtain renormalization group series of record-value length. However, resummation of these series demonstrated [7] that the high order of perturbation theory does not ensure a sufficient accuracy in determining numerical values. For example, the coordinate of the Wilson fixed point specified by a five-loop series for the β function is 5% greater than the known value of high accuracy [11],

$$g^* = 1.7543637(25) \tag{1}$$

and the renormalization group estimate of the critical exponent η differs almost twofold from 1/4 [7]. This situation contrasts sharply with the case of three-dimensional systems [4–6, 12].

Does there exist any method for improving numerical estimates obtained from two-dimensional renormalization group expansions? Below, it will be demonstrated that such a method exists. This method consists in transforming the renormalization group expansions into alternative power series with the coefficients exhibiting a more favorable behavior. The case in point is the implementation of the pseudo-ϵ-expansion technique proposed by B. Nickel (see reference [19] in the paper by Le Guillou and Zinn-Justin [5]). The idea put forward by Nickel is that the coefficient of the linear term in the expansion of the β function should be replaced by a fictitious small parameter τ and that the coordinate of a nontrivial fixed point $g^*$ should be sought in the form of a power series in the parameter τ in order to obtain the τ expansions for the critical exponents. Actually, this technique has already been employed for calculating the critical exponents in two dimensions [5]; however, the relatively short series used in these calculations have made demonstrating the advantages of the method impossible.

## 2. CALCULATION TECHNIQUE

We will operate with a two-dimensional massive theory of the $\lambda\varphi^4$ type normalized to zero external momenta. In this case, the five-loop renormalization group expansions for the β function and the critical exponents γ and η have the following form [7]:

$$\frac{\beta(g)}{2} = -g + g^2 - 0.716173621g^3 + 0.930766443g^4 \\ - 1.58238834g^5 + 3.26042g^6, \tag{2}$$

$$\gamma^{-1} = 1 - \frac{1}{3}g + 0.125023295g^2 - 0.122455138g^3 \\ + 0.164004651g^4 - 0.288554g^5, \tag{3}$$

$$\eta = 0.033966147g^2 - 0.002022555g^3 \\ + 0.011393097g^4 - 0.0137362g^5. \tag{4}$$

Here, the refined value of the five-loop contribution to the inverse exponent $\gamma^{-1}$ is taken from [10]. Let us substitute $-\tau g$ for the first term on the right-hand side of relationship (2) and implement the algorithm described above. As a result, we obtain the expressions

$$g^* = \tau + 0.716173621\tau^2 + 0.095042867\tau^3 \\ + 0.086080396\tau^4 - 0.204139\tau^5, \tag{5}$$

$$\gamma^{-1} = 1 - \frac{1}{3}\tau - 0.113701246\tau^2 + 0.024940678\tau^3 \quad (6)$$
$$- 0.039896059\tau^4 + 0.0645212\tau^5,$$

$$\eta = 0.033966147\tau^2 + 0.046628762\tau^3 \quad (7)$$
$$+ 0.030925471\tau^4 + 0.0256843\tau^5.$$

It can be seen that the power series in $\tau$ for $g^*$ and $\gamma^{-1}$ compare favorably with the renormalization group expansions in the space of physical dimension, because their higher coefficients, even if irregular in sign, are small in magnitude. The smallness of these coefficients allows one to obtain reliable numerical estimates from relationships (5) and (6) without invoking popular summation techniques based on the Borel transform.

This can be easily verified by constructing the Padé approximants [$L/M$] with a parameter $\tau = 1$ for $g^*$ and $\gamma^{-1}$.

## 3. RESULTS AND DISCUSSION

The results obtained for $g^*$ and $\gamma^{-1}$ are presented in Tables 1 and 2, respectively. Since the $\tau$ expansion for the coordinate of the fixed point $g^*$ begins with the linear term, the maximum rank of the approximants $L + M$ in this case is actually equal to 4, whereas for the inverse exponent $\gamma^{-1}$, we have $(L + M)_{max} = 5$. For this reason, the numbers of rows and columns in Table 2 exceed those in Table 1 by unity. The subscripts on the numbers in the tables indicate the coordinates of those poles of the Padé approximants which lie on the real positive $\tau$ semiaxis. The best approximation properties are exhibited by the diagonal Padé approximants [$L/L$] and those close to them which do not have poles at a parameter $\tau > 0$. Therefore, the most reliable estimates of the coordinate $g^*$ should be the numbers 1.751 and 1.837 from Table 1. Averaging over these values, we obtain $g^* = 1.794$, which differs from the exact value (1) by only 2%. As can be seen from Table 1, it is this five-loop approximation that gives such a good estimate; almost all the Padé approximants have "dangerous" poles in lower orders, which leads to a considerable scatter in the numerical values. It seems likely that it is this scatter that led to pessimism in earlier calculations with four-loop series [5].

A similar situation occurs when calculating the critical exponent $\gamma$. It can be seen from Table 2 that, in this case also, reliable estimates are obtained only in the five-loop approximation. Indeed, the numbers specified by the main working approximants [2/3] and [3/2], as well as by the approximant [4/1], almost coincide with each other and are close to the exact value $\gamma = 1.75$. In contrast, the approximants [2/2] and [1/3], which correspond to the four-loop approximation, have dangerous poles. It is worth noting that the pole of the second approximant is located in the vicinity of the physical value $\tau = 1$, which substantially affects the result obtained.

Now, we calculate the Fisher exponent $\eta$. From the comparison of series (4) and (7), we can conclude that, in this case, the pseudo-$\epsilon$-expansion does not lead to any advantages. Furthermore, upon changing over to the expansion in terms of the parameter $\tau$, we obtain not an alternating series but a series of constant signs in which the coefficients are approximately equal in magnitude. By adding four terms of this series at $\tau = 1$, we obtain the Fisher exponent $\eta = 0.137$, whereas the only working Padé approximant [2/2] (all the other approximants have dangerous poles) gives $\eta = 0.0565$. Both of these estimates differ significantly from the exact value $\eta = 0.25$, as result of the processing of the renormalization group expansion (4) using the Padé–Borel–Leroy technique, i.e., $\eta = 0.146$ [7].

We made an attempt to improve the situation. For this purpose, instead of series (7) for the "small" exponent $\eta$, we processed series for the "large" exponents $\nu$ and $\eta^{(2)}$, which are related to the exponent $\eta$ through the

**Table 1.** Coordinate of the Wilson fixed point $g^*$ according to calculations from pseudo-$\epsilon$-expansion (5) with the use of the Padé approximants [$L/M$]

| $M/L$ | 1 | 2 | 3 | 4 | 5 |
|---|---|---|---|---|---|
| 0 | 1.000 | 1.716 | 1.811 | 1.897 | 1.693 |
| 1 | $3.523_{1.4}$ | $1.826_{7.5}$ | $2.724_{1.1}$ | 1.837 | |
| 2 | 1.425 | $1.918_{3.0}$ | $1.850_{6.1}$ | | |
| 3 | $2.601_{1.4}$ | 1.751 | | | |
| 4 | 1.194 | | | | |

Note: In Tables 1–4, the subscripts on the numbers indicate the coordinates of the dangerous poles of the corresponding approximants, i.e., the poles lying on the real positive semiaxis.

**Table 2.** Critical exponent $\gamma$ calculated by the Padé summation of expansion (6) for $\gamma^{-1}$

| $M/L$ | 0 | 1 | 2 | 3 | 4 | 5 |
|---|---|---|---|---|---|---|
| 0 | 1.000 | 1.500 | 1.808 | 1.730 | 1.859 | 1.660 |
| 1 | 1.333 | $2.024_{2.9}$ | 1.744 | 1.778 | 1.777 | |
| 2 | 1.558 | 1.702 | $1.800_{5.2}$ | 1.777 | | |
| 3 | 1.646 | $6.871_{1.1}$ | 1.772 | | | |
| 4 | 1.732 | 1.718 | | | | |
| 5 | $1.714_{6.1}$ | | | | | |

**Table 3.** Pade triangle for the critical exponent $\nu$ calculated by the summation of pseudo-$\epsilon$-expansion (9) for $\nu^{-1}$

| M/L | 0 | 1 | 2 | 3 | 4 | 5 |
|---|---|---|---|---|---|---|
| 0 | 0.500 | 0.750 | 0.933 | 0.920 | 1.005 | 0.898 |
| 1 | 0.667 | $1.107_{2.6}$ | 0.921 | 0.931 | 0.955 | |
| 2 | 0.788 | 0.901 | $0.971_{3.6}$ | $0.959_{5.2}$ | | |
| 3 | 0.846 | $2.999_{1.1}$ | 0.959 | | | |
| 4 | 0.903 | 0.907 | | | | |
| 5 | 0.907 | | | | | |

**Table 4.** Pade triangle for the universal ratio $R_6$ specified by pseudo-$\epsilon$-expansion (12)

| M/L | 1 | 2 | 3 | 4 |
|---|---|---|---|---|
| 0 | 4.000 | 2.364 | 3.587 | 1.837 |
| 1 | 2.839 | 3.064 | 2.867 | |
| 2 | $3.148_{4.5}$ | 2.940 | | |
| 3 | 2.621 | | | |

standard expression. To accomplish this, we find the following expansions in terms of $\tau$ for $\nu$ and $\nu^{-1}$:

$$\nu = \frac{\gamma}{2-\eta} = \frac{1}{2} + \frac{1}{6}\tau + 0.1208977\tau^2 + 0.0584363\tau^3 \quad (8)$$
$$+ 0.0568918\tau^4 + 0.0037987\tau^5,$$

$$\frac{1}{\nu} = 2 - \frac{2}{3}\tau - 0.2613686\tau^2 + 0.0145746\tau^3 \quad (9)$$
$$- 0.0913127\tau^4 + 0.118121\tau^5.$$

It turned out that the first of these expansions is of little use for obtaining numerical estimates: all the Padé approximants generated by this expansion, except for the approximants [5/0] and [0/5], have dangerous poles located in the vicinity of the physical value $\tau = 1$. Therefore, the series for $\nu$ admits only direct summation and also summation of the corresponding inverse series. These operations lead to almost coinciding results, namely, $\nu = 0.907$ and $\nu = 0.898$, which, however, differ from the exact value $\nu = 1$.

The expansion of the inverse exponent $\nu^{-1}$, on the contrary, has a favorable structure for the Padé summation. As follows from the results given in Table 3, all except one of the higher (the fifth order) approximants are free from dangerous poles. Moreover, the approximants [2/3], [3/2], and [4/1] lead to very close values. Nonetheless, assuming that the values of $\gamma = 1.78$ and $\nu = 0.96$, which follow from the results presented in Tables 2 and 3, are the most reliable estimates, we obtain the critical exponent $\eta = 0.156$, which is only scarcely better than the direct estimate $\eta = 0.137$.

Unfortunately, the calculations with the critical exponent $\eta^{(2)} = (2 - \eta)(\gamma^{-1} - 1)$ are also ineffective. Although the pseudo-$\epsilon$-expansion for this exponent can be efficiently summed using the Padé technique, since all the higher approximants do not have poles at $\tau > 0$ and the most symmetric of these approximants give close values of $\eta^{(2)}$ (−0.851, −0.854, −0.837), the final estimate of the exponent differs significantly from the exact value $\eta^{(2)} = -0.75$.

Apart from the critical exponents, some other quantities also take on universal values at a temperature $T \longrightarrow T_c$. In particular, these are the higher effective coupling constants $g_6, g_8, \ldots$, which enter into the equation of state and determine the nonlinear susceptibilities $\chi_{2n}$ of the system (see, for example, [13–17]). These constants can be represented in the form of power series in the renormalized charge $g$. At present, the renormalization group expansion for the constant $g_6$ of the two-dimensional scalar theory $\lambda\varphi^4$ is known in the four-loop approximation [16]:

$$g_6 = \frac{4\pi^2}{81}g^3(1 - 1.125210g + 1.822531g^2 - 3.64849g^3). \quad (10)$$

By substituting expansion (5) into this series, we can readily obtain the pseudo-$\epsilon$-expansion for the sextic effective coupling constant:

$$g_6 = \frac{4\pi^2}{81}(\tau^3 + 1.023311\tau^4 + 0.422991\tau^5 + 0.021201\tau^6). \quad (11)$$

The coefficients of series (11) decrease rapidly in magnitude. However, attempting to sum this series with the use of the Padé approximants leads to the problem of dangerous poles. Only one of the approximants corresponding to the four-loop approximation, namely, the approximant [4/2], is free from dangerous poles; the calculation with this approximant gives the coupling constant $g_6 = 1.122$. This estimate agrees well with the result of the summation of the renormalization group expansion (10) using the Padé–Borel–Leroy technique, i.e., $g_6 = 1.10$ [16].

On the other hand, it is known that the equation of state and the expression for the nonlinear susceptibility $\chi_6$ involve not the vertex $g_6$ itself but the ratio $R_6 = g_6/g_4^2$ [13–17], where $g_4 = g\pi/9$ [16]. Hence, it would be interesting to obtain a power series in $\tau$ for the above ratio. Such a series has the form

$$R_6 = 4\tau(1 - 0.409036\tau + 0.305883\tau^2 - 0.437676\tau^3). \quad (12)$$

The higher coefficients of this expansion do not exhibit a pronounced tendency toward a decrease. Nonetheless, even in this case, the use of the Padé approximants turns out to be efficient. As can be seen from Table 4, only the approximant [1/2] has a pole at $\tau > 0$ and the numbers specified by the working approximants [2/2] and [3/1] are very close to each other. By averaging these values, we obtain $R_6 = 2.90$. This estimate differs by only 1.5% from the results of analyzing the multiloop renormalization group series ($R_6 = 2.94$ [16], $R_6 = 2.95 \pm 0.03$ [18]) and the high-temperature expansions ($R_6 = 2.943 \pm 0.007$ [19]), as well as from recently obtained values of high accuracy ($R_6 = 2.94294$ [11], $R_6 = 2.94238$ [9, 20]).

The computational potential of the pseudo-$\epsilon$-expansion as applied to the ratio $R_6$ is not exhausted by the above estimate. This ratio can be refined by resumming series (12) using the Padé–Borel technique. The calculations demonstrate that the approximants [2/2] and [3/1] constructed for the Borel transform of $R_6$ do not have dangerous poles, and their processing leads to the values $R_6 = 2.970$ and $R_6 = 2.909$, respectively. Averaging these values, we obtain the ratio $R_6 = 2.94$, which coincides with the results of the calculations performed in [11, 16, 19, 20].

In conclusion, we should note that the Ising model is not a unique system for which the higher coefficients of the pseudo-$\epsilon$-expansions are small as compared to the coefficients of the renormalization group series in $\epsilon$ and $g$. A similar feature was revealed recently for the three-dimensional cubic model [21], the three-dimensional chiral model [22], and the two-dimensional *MN* model [10]. This made it possible, in particular, to obtain alternative numerical estimates for the marginal dimensions of the order parameter, i.e., those values of *M* and *N* which separate the regions with different regimes of the critical behavior [10, 21, 22].

## ACKNOWLEDGMENTS

The author would like to thank E.V. Orlov for performing some control calculations.

This work was supported by the Russian Foundation for Basic Research, project no. 04-02-16189.